\documentclass[%
reprint,
superscriptaddress,
showpacs,
nofootinbib,
amsmath,amssymb,
prc,
floatfix ]%
{revtex4-1}

\usepackage{color}
\usepackage{CJK}

\usepackage{graphicx}
\usepackage{dcolumn}
\usepackage{bm}
\usepackage{hyperref}

\allowdisplaybreaks

\begin{document}

\begin{CJK*}{UTF8}{}
\title{Time-dependent generator coordinate method study of fission: mass parameters}
\CJKfamily{gbsn}
\author{Jie Zhao (赵杰)}%
\affiliation{Center for Quantum Computing, Peng Cheng Laboratory, Shenzhen 518055, China}
\author{Tamara Nik\v{s}i\'c}%
\affiliation{Physics Department, Faculty of Science, University of Zagreb, Bijeni\v{c}ka Cesta 32,
        	      Zagreb 10000, Croatia}             
\author{Dario Vretenar}%
\affiliation{Physics Department, Faculty of Science, University of Zagreb, Bijeni\v{c}ka Cesta 32,
              Zagreb 10000, Croatia}
\CJKfamily{gbsn}              
\author{Shan-Gui Zhou (周善贵)}%
 \affiliation{CAS Key Laboratory of Theoretical Physics, Institute of Theoretical Physics, Chinese Academy of Sciences, Beijing 100190, China}
 \affiliation{School of Physical Sciences, University of Chinese Academy of Sciences, Beijing 100049, China}
 \affiliation{Center of Theoretical Nuclear Physics, National Laboratory of Heavy Ion Accelerator, Lanzhou 730000, China}
 \affiliation{Synergetic Innovation Center for Quantum Effects and Application, Hunan Normal University, Changsha 410081, China}

\date{\today}

\begin{abstract}
Collective mass tensors derived in the cranking approximation to the adiabatic time-dependent Hartree-Fock-Bogoliubov (ATDHFB) method are employed in a study of induced fission dynamics. Together with a collective potential determined in deformation-constrained self-consistent mean-field calculations based on nuclear energy density functionals, the mass tensors specify the collective Hamiltonian that governs the time evolution of the nuclear wave function from an initial state at equilibrium deformation, up to scission and the formation of fission fragments. In an illustrative calculation of low-energy induced fission of $^{228}$Th, $^{230}$Th, $^{234}$U, and $^{240}$Pu, we compare the non-perturbative and perturbative cranking ATDHFB mass tensors in the plane of axially-symmetric quadrupole and octupole deformations, as well as the resulting charge yields.
\end{abstract}

\maketitle

\end{CJK*}

\bigskip

\section{Introduction~\label{sec:Introduction}}

Models based on the generator coordinate method (GCM) \cite{RingSchuck} have successfully been applied to studies of both low-energy spectroscopic properties and fission dynamics in a single theoretical framework. The time-dependent version of this method (TDGCM), in particular, can describe the entire process of induced fission from some initial state through a complex time-evolution of collective degrees of freedom, leading up to scission and the emergence of fission fragments \cite{KrappePomorski,Schunck2016_RPP79-116301,Younes2019}. In the Gaussian overlap approximation (GOA) the TDGCM is represented by a local Schr\"odinger equation for the nuclear wave function in the space of collective coordinates. This equation and, therefore, the description of fission dynamics are determined by the collective potential and inertia that are typically computed in a self-consistent mean-field framework based on an energy density functional (EDF) or effective nuclear interaction. For a particular choice of collective degrees of freedom such as, for instance, variables that characterize the elongation, shape and asymmetry of the fissioning nucleus, the collective potential is almost completely (up to the zero-point energy correction) determined by the diagonal matrix elements of the effective Hamiltonian in the non-orthogonal basis of static symmetry-breaking product many-body states. Much more challenging, both conceptually as well as from a computational point of view, is the collective inertia tensor. 

Two methods have been used to derive the collective masses for fission: the GCM+GOA and the adiabatic time-dependent Hartree-Fock-Bogoliubov (ATDHFB). It is well known that the standard GCM+GOA method does not lead to the correct collective mass such as, for example, the bare mass of the nucleus in the simple case of pure translation \cite{RingSchuck,Younes2019}. The proper collective mass could only be obtained if, in addition to the collective coordinates, also the corresponding conjugate momenta were taken into account in the GCM. However, this means that one has to double the dimension of the collective space, and this is never done in practical applications to fission. The alternative has been to use ATDHFB collective masses, but even in that case the exact expression for the collective mass requires the inversion of the full linear response matrix. For this reason non-perturbative and perturbative cranking approximations to the ATDHFB masses have been derived \cite{Yuldashbaeva1999_PLB461-1,Baran2011_PRC84-054321} and applied to fission studies. 

In the perturbative cranking approximation the contribution from time-odd mean fields is neglected, and derivatives of the single-nucleon and pairing densities with respect to collective coordinates are calculated perturbatively.  The non-perturbative cranking ATDHFB collective mass tensor can be computed by explicit numerical evaluation of the derivatives with respect to collective coordinates. Detailed studies of spontaneous fission half-lives with the collective mass tensors calculated using the ATDHFB method both in the
perturbative and non-perturbative cranking approximations \cite{Sadhukhan2013_PRC88-064314,Zhao2015_PRC92-064315} have shown that the structural properties of the collective mass crucially determine the dynamics of spontaneous fission. In a recent comparative analysis of non-perturbative collective inertias for fission \cite{Giuliani2018_PLB787-134} it has been shown that non-perturbative methods based on both the GCM+GOA and ATDHFB predict very similar collective masses with a much more complex structure that those obtained in the perturbative approach. In both the non-perturbative and perturbative calculations the ATDHFB masses were larger than the corresponding GCM+GOA masses by a factor $\approx 1.5$, almost constant over the whole range of axial quadrupole deformation extending to the region where two separate fragments emerge.   

In all applications of the TDGCM framework to induced fission dynamics, however, only perturbative cranking ATDHFB collective masses have been employed so far \cite{Goutte2005_PRC71-024316,Regnier2016_PRC93-054611,Zdeb2017_PRC95-054608,Regnier2018_CPC225-180,Regnier2019_PRC99-024611,Tao2017_PRC96-024319,Zhao2019_PRC99-014618,Zhao2019_PRC99-054613}. The goal of this study is to explore differences between non-perturbative and perturbative ATDHFB collective masses when used in TDGCM+GOA modeling of low-energy induced fission dynamics. The theoretical framework and methods are briefly reviewed in Sec.~\ref{sec:model}.
The details of the calculation, the results for deformation energy surfaces, 
collective masses, as well as the resulting charge yield distributions for induced fission of 
$^{228}$Th, $^{230}$Th, $^{234}$U, and $^{240}$Pu 
are described and discussed in Sec.~\ref{sec:results}.
Sec.~\ref{sec:summary} contains a short summary of the principal results.

\section{\label{sec:model}Theoretical framework}
The particular implementation of the TDGCM+GOA collective Hamiltonian used in the present study 
is described in Refs.~\cite{Tao2017_PRC96-024319,Zhao2019_PRC99-014618,Zhao2019_PRC99-054613}, 
and the computer code employed for modeling the time evolution of the fissioning nucleus is FELIX (version 2.0) \cite{Regnier2018_CPC225-180}. For completeness here we include a brief outline of the model and 
discuss the basic approximations. 

In the TDGCM+GOA framework induced fission is described as a slow adiabatic process determined by a small number of  collective degrees of freedom. Nonadiabatic effects arising from the coupling between collective and intrinsic degrees 
of freedom are not taken into account. Fission dynamics is thus governed by a local, time-dependent Schr\"odinger-like equation 
in the space of collective coordinates $\bm{q}$:
\begin{equation}  
i\hbar \frac{\partial g(\bm{q},t)}{\partial t} = \hat{H}_{\rm coll} (\bm{q}) g(\bm{q},t) , 
\label{eq:TDGCM}
\end{equation}
where $g(\bm{q},t)$ is the complex wave function of the collective variables $\bm{q}$ and time $t$. 
For simplicity we assume axial symmetry with respect to the axis along which the two fragments eventually separate, and consider the two-dimensional (2D) collective space of deformation parameters: 
quadrupole $\beta_{2}$ and octupole $\beta_{3}$.   
The collective Hamiltonian $\hat{H}_{\rm coll} (\bm{q})$ thus reads
\begin{align}
\hat{H}_{\rm coll} (\beta_2,\beta_3) &= - {\hbar^2 \over 2} \times \nonumber \\ 
&\sum_{ij =2,3} {\partial \over \partial \beta_i} B_{ij}(\beta_2,\beta_3) {\partial \over \partial \beta_j} + V(\beta_2,\beta_3),
\label{eq:Hcoll2}
\end{align}
where $B_{ij}(\beta_2,\beta_3)$ and $V(\beta_2,\beta_3)$ denote the inertia tensor and collective potential, respectively.  
The inertia tensor is the inverse of the mass tensor, that is, $B_{ij}(\beta_2,\beta_3) =(\mathcal{M}^{-1})_{ij}$.
The adiabatic time-dependent Hartree-Fock-Bogoliubov (ATDHFB) method is applied in both the non-perturbative and perturbative cranking approximations to the calculation of the mass tensor.
In the cranking approximation the mass tensor takes the form \cite{Baran2011_PRC84-054321} 
\begin{equation}
\label{eq:npmass}
\mathcal{M}_{ij}^{C} = {\hbar^2 \over 2 \dot{q}_i \dot{q}_j}
    \sum_{\mu\nu} {F^{i*}_{\mu\nu}F^{j}_{\mu\nu} + F^{i}_{\mu\nu}F^{j*}_{\mu\nu}
    \over E_{\mu} + E_{\nu}},
\end{equation}
where
\begin{equation}
\label{eq:fmatrix}
{F^{i} \over \dot{q}_{i}}  
  =  U^\dagger {\partial\rho \over \partial q_{i}} V^* 
    + U^\dagger {\partial\kappa \over \partial q_{i}} U^*
    - V^\dagger {\partial\rho^* \over \partial q_{i}} U^*
    - V^\dagger {\partial\kappa^* \over \partial q_{i}} V^*\;.
\end{equation}
$U$ and $V$ are the self-consistent Bogoliubov matrices, and $\rho$ and $\kappa$ are 
the corresponding particle and pairing density matrices, respectively.
The derivatives of the densities are calculated using the Lagrange three-point formula for 
unequally spaced points~\cite{Yuldashbaeva1999_PLB461-1,Baran2011_PRC84-054321}.
The cranking expression Eq.~(\ref{eq:fmatrix}) can be further simplified in a perturbative 
approach~\cite{Brack1972_RMP44-320,Nilsson1969_NPA131-1,Girod1979_NPA330-40,Bes1961_NP28-42,
Sobiczewski1969_NPA131-67}, and this leads to the perturbative cranking mass tensor 
\begin{equation}
\label{eq:pmass}
\mathcal{M}^{Cp} = \hbar^2 {\it M}_{(1)}^{-1} {\it M}_{(3)} {\it M}_{(1)}^{-1}, 
\end{equation}
where 
\begin{equation}
\label{eq:mmatrix}
\left[ {\it M}_{(k)} \right]_{ij} = \sum_{\mu\nu} 
    {\left\langle 0 \left| \hat{Q}_i \right| \mu\nu \right\rangle
     \left\langle \mu\nu \left| \hat{Q}_j \right| 0 \right\rangle
     \over (E_\mu + E_\nu)^k}.
\end{equation}
$|\mu\nu\rangle$ are two-quasiparticle states and $E_\mu$, $E_\nu$ denote the corresponding quasiparticle energies. 
Details of the derivation of the cranking formulas for the mass tensor can be found in 
Ref.~\cite{Baran2011_PRC84-054321}. 

The input for the calculation of the collective mass, that is, the single-quasiparticle states, energies, and occupation factors are calculated in a self-consistent mean-field approach based on nuclear energy density 
functionals. The map of the energy surface as function of the quadrupole and octupole deformations is obtained by imposing constraints on the corresponding mass moments:
\begin{equation}
\label{eq:multipole-moments}
\hat{Q}_2 = 2z^2 - r_\perp^2 \quad \textnormal{and} \quad \hat{Q}_3 = 2z^3 - 3z r_\perp^2.
\end{equation}
The deformation parameters $\beta_2$ and $\beta_3$ are determined using the following relations:
\begin{equation}
\beta_2 = \frac{\sqrt{5\pi}}{3AR_0^2} \langle \hat{Q}_2 \rangle \quad \textnormal{and} \quad 
\beta_3 = \frac{\sqrt{7\pi}}{3AR_0^3} \langle \hat{Q}_3 \rangle,
\end{equation}
with $R_0=r_0A^{1/3}$ and $r_0=1.2$ fm.
The collective potential $V(\beta_2,\beta_3)$ is obtained by subtracting the vibrational zero-point energy (ZPE) from the total
mean-field energy  \cite{Staszczak2013_PRC87-024320}
\begin{equation}
\label{eq:zpe}
E_{\rm ZPE} = {1\over4} {\rm Tr} \left[ {\it M}_{(2)}^{-1} {\it M}_{(1)} \right],
\end{equation}
where the ${\it M}_{(k)}$ are given by Eq.~(\ref{eq:mmatrix}).

The collective space is divided into an inner region with a single nuclear density distribution, 
and an external region that contains two separated fission fragments. 
The set of configurations that divides the inner and external regions defines the scission 
hyper-surface.  The flux of the probability current through this
hyper-surface provides a measure of the probability of observing a given pair of fragments at time $t$.
Each infinitesimal surface element is associated with a given pair of fragments $(A_L, A_H)$, where $A_L$ and $A_H$ denote the 
lighter and heavier fragments, respectively.
The integrated flux $F(\xi,t)$ for a given surface element $\xi$ is defined as \cite{Regnier2018_CPC225-180}
\begin{equation}
F(\xi,t) = \int_{t_0}^{t} dt^\prime \int_{\{ \beta_2, \beta_3 \}\in \xi} \bm{J}(\beta_2,\beta_3,t^\prime) \cdot d\bm{S}, 
\label{eq:flux}
\end{equation}
where $\bm{J}(\beta_2,\beta_3,t)$ is the current
\begin{align}
\label{eq:current}
J_k(\beta_2,\beta_3,t) &= \hbar \sum_{l\in \{2,3\}}{
B_{kl}(\beta_2,\beta_3) {\mathrm{Im}}\left(g^* \frac{\partial g}{\partial \beta_l} \right)}.
\end{align}
The yield for the fission fragment with mass $A$ is defined by 
\begin{equation}
Y(A) \propto \sum_{\xi \in \mathcal{A}} \lim_{t \rightarrow \infty} F(\xi,t).
\end{equation}
The set $\mathcal{A}(\xi)$ contains all elements belonging to the scission hyper-surface such that one of the fragments has mass number $A$.

In the present study mean-field energy surfaces are calculated with the multidimensionally constrained relativistic mean-field (MDC-RMF) model \cite{Lu2012_PRC85-01301R,Lu2014_PRC89-014323,Zhou2016_PS91-063008,Zhao2016_PRC93-044315},  
using the point-coupling relativistic energy density functional
DD-PC1~\cite{Niksic2008_PRC78-034318}. Pairing correlations are taken into account in the BCS approximation with a separable
pairing force of finite range \cite{Tian2009_PLB676-44}:
\begin{equation}
V(\mathbf{r}_1,\mathbf{r}_2,\mathbf{r}_1^\prime,\mathbf{r}_2^\prime) = G_0 ~\delta(\mathbf{R}-
\mathbf{R}^\prime) P (\mathbf{r}) P(\mathbf{r}^\prime) \frac{1}{2} \left(1-P^\sigma\right),
\label{pairing}
\end{equation}
where $\mathbf{R} = (\mathbf{r}_1+\mathbf{r}_2)/2$ and $\mathbf{r}=\mathbf{r}_1- \mathbf{r}_2$
denote the center-of-mass and the relative coordinates, respectively. $P(\mathbf{r})$ reads 
\begin{equation}
P(\mathbf{r})=\frac{1}{\left(4\pi a^2\right)^{3/2}} e^{-\mathbf{r}^2/4a^2}.
\end{equation}
The parameters of the interaction were originally 
adjusted to reproduce the density dependence of the pairing gap in nuclear matter at the
Fermi surface computed with the D1S parameterization of the Gogny force~\cite{Berger1991_CPC63-365}.
To reproduce the empirical pairing gaps in the mass region considered in the present study, the strength parameters 
of the pairing force have been increased 
with respect to the original values by the following factors: $G_{n}/G_{0}=1.12$ and $G_{p}/G_{0}=1.08$ for 
neutrons and protons, respectively.

The fission process is described by 
 the time evolution of an initial wave packet $g(\bm{q},t=0)$ ($\bm{q} \equiv \{\beta_2,\beta_3\}$), built 
as a Gaussian superposition of the quasi-bound states $g_k$, 
\begin{equation}
g(\bm{q},t=0) = \sum_{k} \exp\left( { (E_k - \bar{E} )^{2} \over 2\sigma^{2} } \right) g_{k}(\bm{q}),
\label{eq:initial-state}
\end{equation}
where the value of the parameter $\sigma$ is set to 0.5 MeV. The collective states $\{ g_{k}(\bm{q}) \}$ 
are solutions of the stationary eigenvalue equation in which the original collective potential $V(\bm{q})$ is replaced by a 
new potential $V^{\prime} (\bm{q})$ that is obtained by extrapolating the inner potential barrier with a quadratic form. 
The mean energy $\bar{E}$ in Eq.~(\ref{eq:initial-state}) is then adjusted iteratively in 
such a way that $\langle g(t=0)| \hat{H}_{\rm coll} | g(t=0) \rangle = E_{\rm coll}^{*}$, and this 
average energy $E_{\rm coll}^{*}$ is chosen $\approx 1$ MeV above the fission barrier.
The TDGCM+GOA Hamiltonian of Eq.~(\ref{eq:Hcoll2}), with the original collective potential 
$V(\bm{q})$, propagates the initial wave packet in time.

The time propagation is modeled using the TDGCM+GOA computer 
code FELIX (version 2.0)~\cite{Regnier2018_CPC225-180}.
The time step is $\delta t=5\times 10^{-4}$ zs (1 zs $= 10^{-21}$ s), and the charge and mass 
distributions are calculated after $10^{5}$ time steps, which correspond to 50 zs.
As in our recent calculations of Refs.~\cite{Tao2017_PRC96-024319,Zhao2019_PRC99-014618,Zhao2019_PRC99-054613}, 
the parameters of the additional imaginary absorption potential that takes into account the escape 
of the collective wave packet  in the domain outside the region of calculation \cite{Regnier2018_CPC225-180} are: 
the absorption rate $r=20\times 10^{22}$ s$^{-1}$ and the width of the absorption band $w=6.0$.
The charge yields are obtained by convoluting the raw flux with a Gaussian function of the number of particles  \cite{Regnier2016_PRC93-054611,Zhao2019_PRC99-014618}, with a width of 1.6 units.

\section{\label{sec:results}Results and discussion}
To illustrate the effect of a particular choice of collective inertia on the fragment distribution, in this section we discuss 
results for the process of induced fission of $^{228}$Th, $^{230}$Th, $^{234}$U, and $^{240}$Pu. In the first step a large scale  
MDC-RMF calculation is performed to generate the potential energy surface, 
single-nucleon wave functions and occupation factors in the $(\beta_2,\beta_3)$ plane.  
The range for the collective variable $\beta_2$ is $0 \le \beta_2 \le 7$ with   
a step $\Delta \beta_2 = 0.04$, while the collective variable $\beta_3$ is considered in the interval $0 \le \beta_3 \le 3.5$ with a step $\Delta \beta_3 =0.05$. The relativistic energy density functional DD-PC1 is used in the particle-hole channel, while 
particle-particle correlations are described by the separable pairing force (\ref{pairing}) in the BCS approximation.

The deformation energy surface is determined in a self-consistent calculation with constraints on the mass multipole
moments $Q_{2}$ and $Q_{3}$  Eq.~(\ref{eq:multipole-moments}), by employing the augmented Lagrangian method~\cite{Staszczak2010_EPJA46-85}.
The mean-field equations are solved by expanding the nucleon Dirac spinors in the axially deformed harmonic oscillator (ADHO)
basis with $N_f=20$ oscillator shells. Ref.~\cite{Lu2014_PRC89-014323} details the multidimensionally-constrained relativistic mean-field model.

\begin{figure}
 \includegraphics[width=0.48\textwidth]{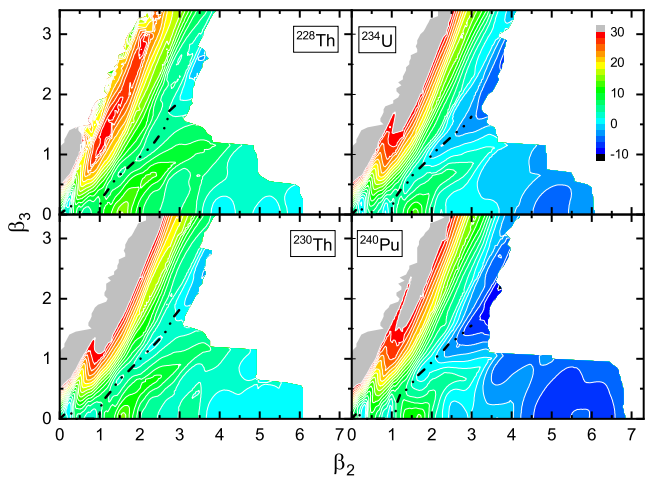}
\caption{(Color online)~\label{fig:Th_Pu_PES}%
Axially-symmetric quadrupole-octupole collective potentials in the 
$\beta_{2} - \beta_{3}$ plane for $^{228}$Th, $^{230}$Th, $^{234}$U, and $^{240}$Pu.
In each panel the energies are normalized with respect to the corresponding value at the equilibrium minimum.
The contours join points on the surface with the same energy, 
and the separation between neighbouring contours is 2 MeV. The dot-dashed curve is the static, lowest-energy fission path.}
\end{figure}

Figure \ref{fig:Th_Pu_PES} displays the resulting quadrupole- and octupole-constrained collective potential 
surfaces of $^{228}$Th, $^{230}$Th, $^{234}$U, and $^{240}$Pu. The vibrational zero-point energies have been subtracted from the total mean-field energies. Only the points in the collective space that belong to the inner
region with a single nuclear density distribution are included in the plots. The scission contour that divides the inner and external regions is determined by the Gaussian neck operator $\displaystyle \hat{Q}_{N}=\exp[-(z-z_{N})^{2} / a_{N}^{2}]$, 
where $a_{N}=1$ fm and $z_{N}$ is the position of the neck~\cite{Younes2009_PRC80-054313}.
We define the pre-scission domain by $\langle \hat{Q}_{N} \rangle>3$, and consider the frontier of this domain as the scission contour.
For $^{228}$Th, $^{230}$Th, and $^{234}$U the scission line starts from an elongated symmetric point with $\beta_2 \approx 6$, while
for $^{240}$Pu this value is somewhat larger. As the asymmetry $\beta_3$ increases, the
scission profile evolves to smaller $\beta_2$ deformations for all four nuclei. The ridge separating the asymmetric and symmetric
fission valleys is more pronounced for $^{228}$Th and $^{230}$Th, while it is lower for $^{234}$U and $^{240}$Pu. 
The dot-dashed curves correspond to the static, lowest-energy fission paths.

%

\begin{figure}
 \includegraphics[width=0.48\textwidth]{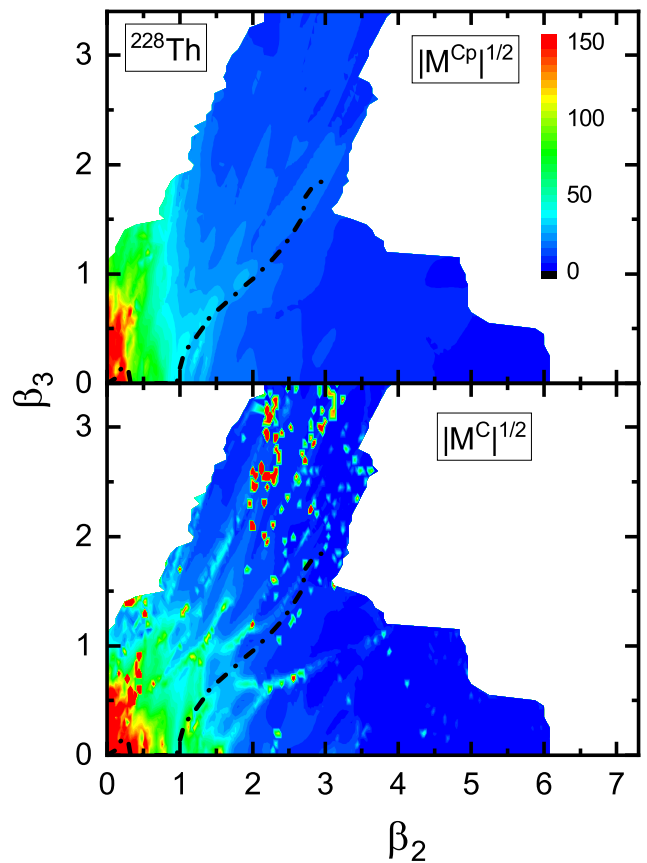}
\caption{(Color online)~\label{fig:Th228_Mass}%
Square-root determinants of the perturbative-cranking mass tensor $|\mathcal{M}^{Cp}|^{1/2}$, 
and nonperturbative-cranking mass tensor $|\mathcal{M}^{C}|^{1/2}$ (in $\hbar^2$ MeV$^{-1}$)
of $^{228}$Th in the $(\beta_{2}, \beta_{3})$ plane. The dot-dashed curve is the static, lowest-energy fission path.
}
\end{figure}

For the two dimensional quadrupole-octupole collective space $\{\beta_2,\beta_3 \}$ the mass tensor is determined by three independent components: $\mathcal{M}_{22}$, $\mathcal{M}_{23}$ and $\mathcal{M}_{33}$. In Fig. \ref{fig:Th228_Mass} 
we plot the square-root determinants 
$|\mathcal{M}|^{1/2} =\left(\mathcal{M}_{22}\mathcal{M}_{33} - \mathcal{M}_{23}^2 \right)^{1/2}$ for $^{228}$Th. The upper panel displays the mass tensors calculated using the perturbative cranking formula Eq.~(\ref{eq:pmass}), 
while the one determined in the non-perturbative cranking method of Eq.~(\ref{eq:npmass}) is shown in the lower panel. 
Just as in Fig. \ref{fig:Th_Pu_PES}, only points that belong to the inner region are included in 
the plot and the dot-dashed curves denote the static fission path. 
The general pattern is similar for all four nuclei considered in the present study and, in particular, one notices that in the non-perturbative approach the values 
of $|\mathcal{M}|^{1/2}$ are enhanced at relatively small deformations, and characterized by isolated peaks in the region of large 
octupole deformations $\beta_3$.  Note, however, that these peaks
are located far outside the asymmetric fission valley. The increase of the 
collective mass in the region $\beta_3 \approx 0$ should weaken the current in that 
region\footnote{The inertia tensor is defined as the inverse of the mass tensor ($B_{ij}(\beta_2,\beta_3) =(\mathcal{M}^{-1})_{ij}$), and the current Eq.~(\ref{eq:current}) is proportional to the collective inertia.}, thus generally reducing the fragment distribution for symmetric fission.




%

\begin{figure}
 \includegraphics[width=0.48\textwidth]{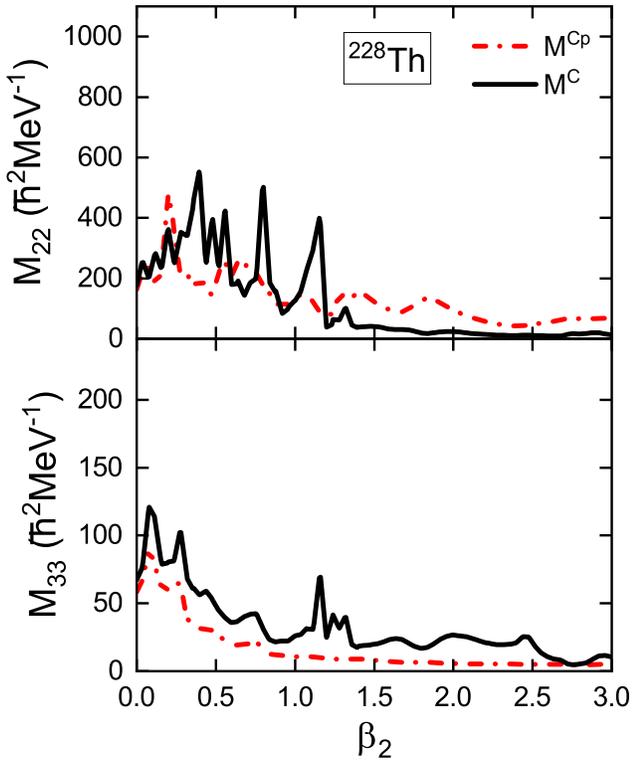}
\caption{(Color online)~\label{fig:Th228_StaticPath_Mass}%
The $\mathcal{M}_{22}$ (upper panel) and $\mathcal{M}_{33}$ (lower panel) components of the mass tensor of $^{228}$Th, 
as function of the quadrupole deformation $\beta_{2}$ along the static fission path.}
\end{figure}

\begin{figure}
 \includegraphics[width=0.48\textwidth]{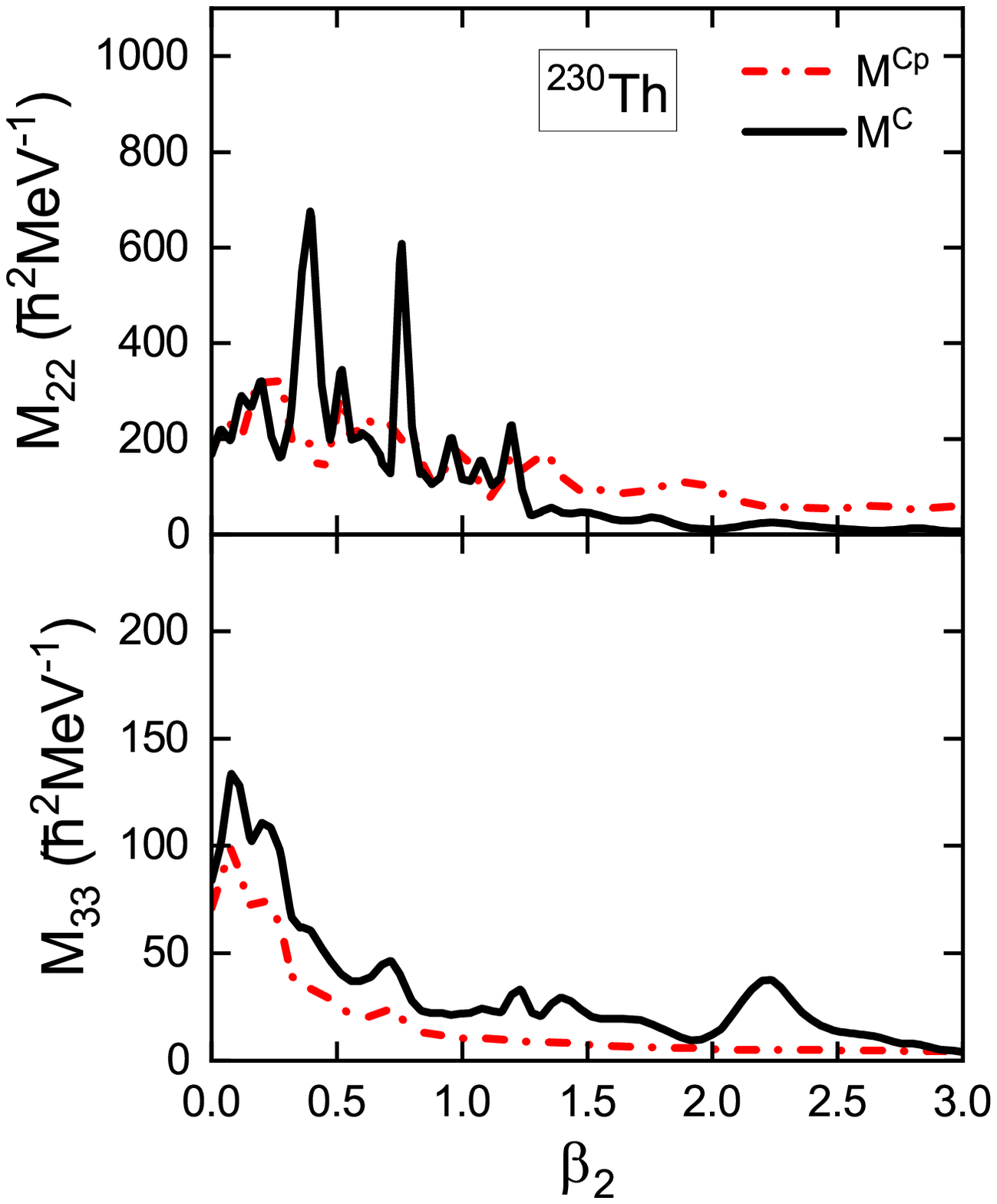}
\caption{(Color online)~\label{fig:Th230_StaticPath_Mass}%
Same as in the caption Fig.~\ref{fig:Th228_StaticPath_Mass} but for $^{230}$Th.}
\end{figure}

\begin{figure}
 \includegraphics[width=0.48\textwidth]{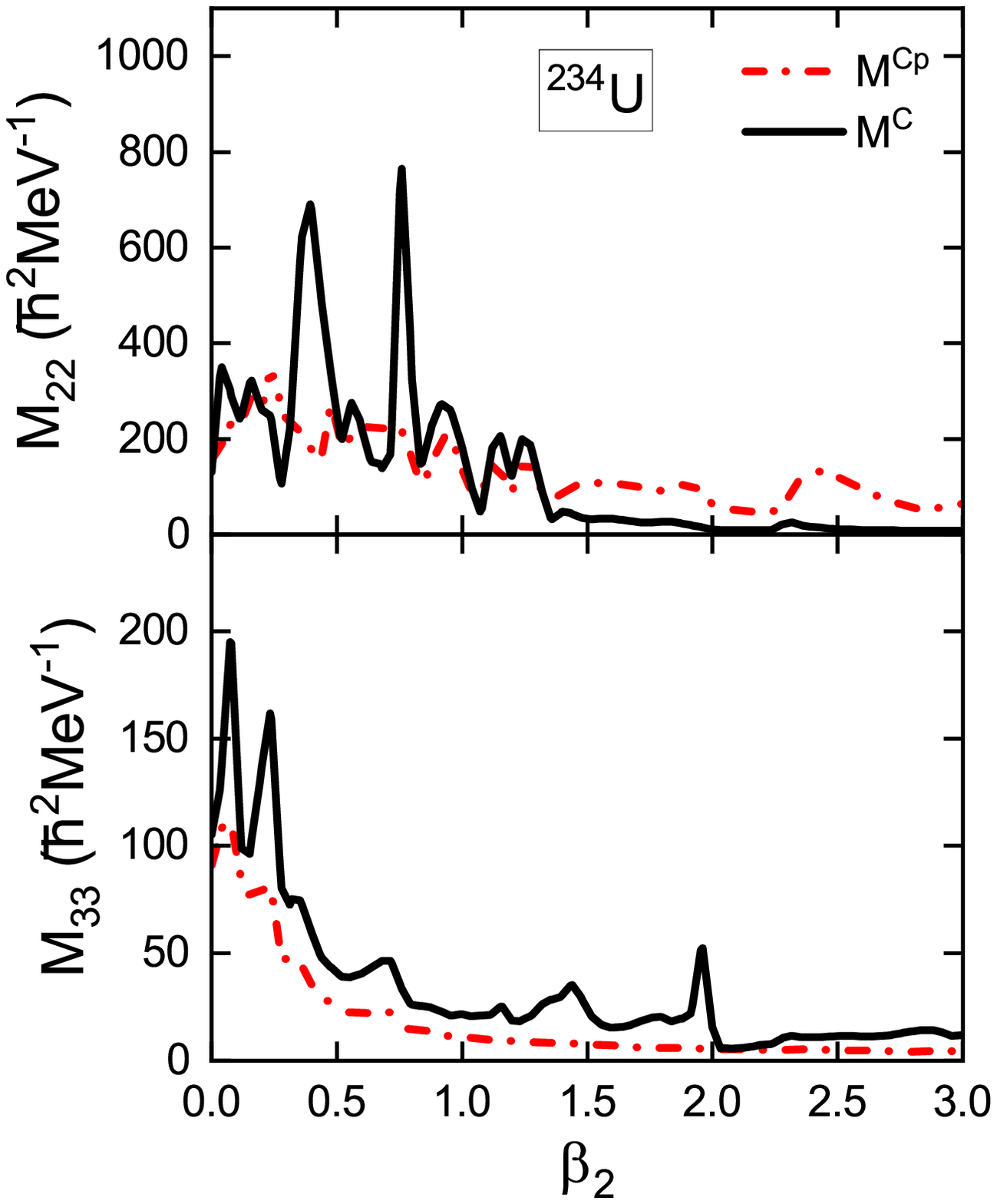}
\caption{(Color online)~\label{fig:U234_StaticPath_Mass}%
Same as in the caption Fig.~\ref{fig:Th228_StaticPath_Mass} but for $^{234}$U.}
\end{figure}

\begin{figure}
 \includegraphics[width=0.48\textwidth]{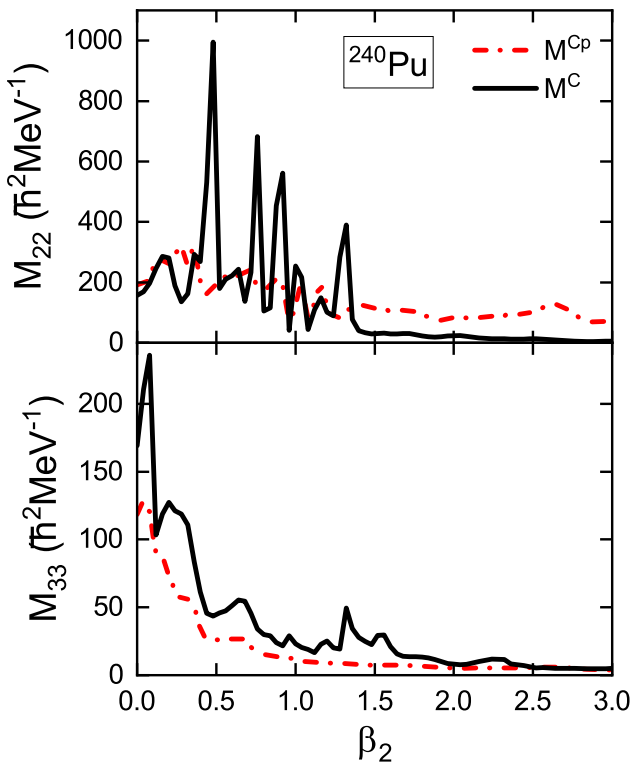}
\caption{(Color online)~\label{fig:Pu240_StaticPath_Mass}%
Same as in the caption Fig.~\ref{fig:Th228_StaticPath_Mass} but for $^{240}$Pu.}
\end{figure}

To illustrate in more detail the differences between the perturbative and nonperturbative cranking mass parameters, in Figs.~\ref{fig:Th228_StaticPath_Mass}-\ref{fig:Pu240_StaticPath_Mass} we plot the diagonal components $\mathcal{M}_{22}$ and $\mathcal{M}_{33}$ of the mass tensor, calculated along the static fission paths for $^{228}$Th, $^{230}$Th, $^{234}$U and $^{240}$Pu, as functions of the quadrupole collective coordinate. Both components calculated using the perturbative cranking formula display a gradual decrease with quadrupole deformation along the static fission path, and we note the oscillations of  $\mathcal{M}_{22}$ especially at smaller deformations. The non-perturbative mass parameters, in particular $\mathcal{M}_{22}$, exhibit sharp peaks in the region $\beta_2 \le 1.5$. The spikes occur because of single-particle level crossings near the Fermi surface, characterized by sudden changes of the occupation factors of single-particle configurations \cite{Baran2011_PRC84-054321, Sadhukhan2013_PRC88-064314}.
For large quadrupole deformations $\beta_2 > 1.5$ both perturbative and nonperturbative mass parameters decrease more smoothly along the static path. It is interesting to note that the perturbative $\mathcal{M}_{22}$ is generally larger than the corresponding non-perturbative mass parameter, while the opposite trend is observed for the $\mathcal{M}_{33}$ component.

\begin{figure*}
 \includegraphics[width=0.75\textwidth]{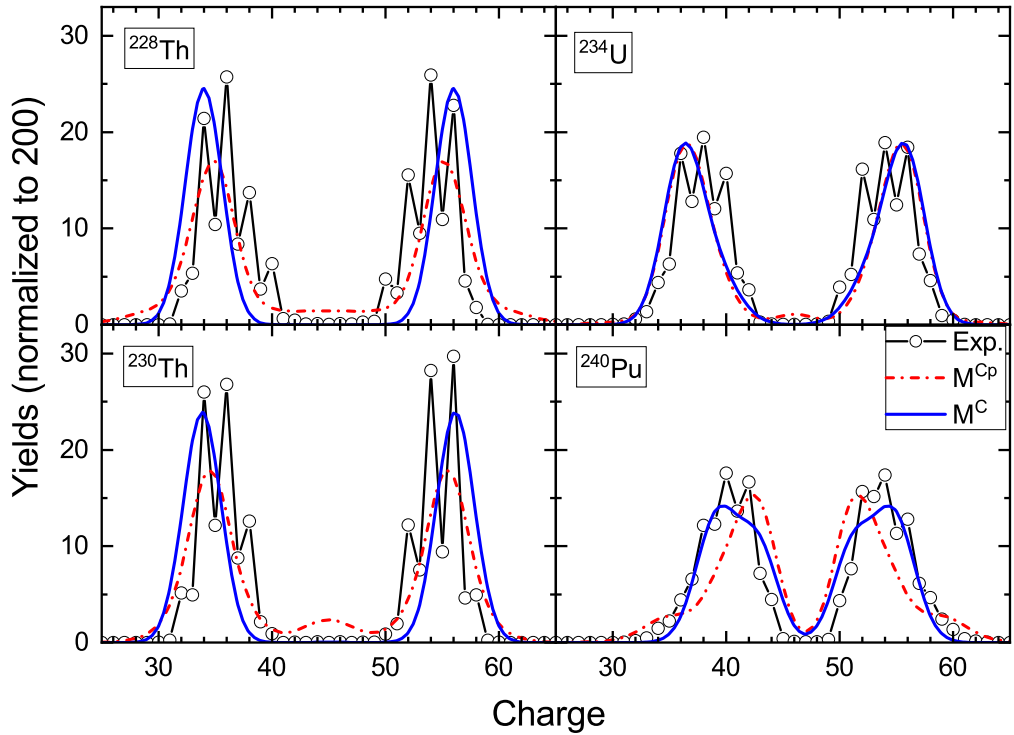}
\caption{(Color online)~\label{fig:Th_Pu_Zyields}%
Charge yields for induced fission in $^{228}$Th, $^{230}$Th, $^{234}$U, and $^{240}$Pu. 
The perturbative ($\mathcal{M}^{Cp}$) and nonperturbative ($\mathcal{M}^{C}$) cranking 
inertia tensors are used in the TDGCM+GOA calculation.
The experimental thermal neutron induced fission charge yields are from Ref.~\cite{NNDC}. 
}
\end{figure*}

In the second step we calculate the charge yields for induced fission of the four nuclei. The initial wave packet is given by Eq.~(\ref{eq:initial-state}) so that the average energy is 1 MeV above the fission barrier, and the time evolution is governed by the collective Hamiltonian (\ref{eq:Hcoll2}). Both perturbative and non-perturbative cranking collective inertia tensors are used to evolve the collective wave packet across the potential energy surface in the $\beta_{2} - \beta_{3}$ plane, and the flux through the scission contour determines the fission yields as described in the previous section. Figure \ref{fig:Th_Pu_Zyields} displays the resulting charge yields for induced fission of $^{228}$Th, $^{230}$Th, $^{234}$U, and $^{240}$Pu. The model obviously cannot describe the odd-even staggering of the experimental charge yields, but otherwise reproduces the empirical distributions. In general we notice a reduction of symmetric yields when the non-perturbative cranking collective inertia are used, thus bringing the results in better agreement with data. This is due to the increase of the collective mass in the region of small octupole deformations and the resulting reduction of the flux for symmetric fission. The effect is very weak in $^{234}$U but somewhat more pronounced for the other three nuclei. 

\section{\label{sec:summary}Summary}

Non-perturbative cranking ATDHFB collective masses have been used for the first time in the TDGCM+GOA description of induced fission dynamics. The mass tensor determines the adiabatic collective motion of the fissioning nucleus governed by the Schr\" odinger equation for the nuclear wave function in the space of deformation parameters. In an illustrative calculation of low-energy induced fission of four actinide nuclei, we have compared the non-perturbative and perturbative ATDHFB mass tensors in the plane of axially-symmetric quadrupole and octupole deformations, as well as the resulting charge yields. As noted in previous studies, the structure of non-perturbative collective masses is much more complex due to changes in the intrinsic shell structure across the deformation energy surface, and it is characterized by pronounced isolated peaks located at single-particle level crossings near the Fermi surface. In the present study we have been able to use both  non-perturbative and perturbative masses in modeling the time-evolution of an initial collective wave packet across the scission contour to the region in which separate fragments emerge. It has been shown that the choice of the collective mass affects the predicted fragment distribution. In the example explored here, the choice of non-perturbative cranking collective mass leads to a reduction of symmetric charge yields and, generally, to a better agreement with data. This result motivates further studies and applications of full cranking ATDHFB masses to  fission dynamics by considering additional collective degrees of freedom such as non-axial shape deformations and dynamical pairing.

\bigskip


\bigskip
\acknowledgements
This work has been supported by the Inter-Governmental S\&T Cooperation Project between China and Croatia. 
It has also been supported in part by the QuantiXLie Centre of Excellence, a project co-financed by the Croatian Government and European Union through the European Regional Development Fund - the Competitiveness and Cohesion Operational Programme (KK.01.1.1.01)
and by the Croatian Science Foundation under the project Uncertainty quantification within the nuclear energy density 
framework (IP-2018-01-5987).
J.Z. acknowledges support by the National Natural Science Foundation of China under Grant No. 11790325.
S.G.Z. has been supported by 
the National Key R\&D Program of China (Grant No. 2018YFA0404402), 
the National Natural Science Foundation of China (Grants 
No. 11525524, No. 11621131001, No. 11947302, and No. 11961141004),
the Key Research Program of Frontier Sciences of Chinese Academy of Sciences (Grant No. QYZDB-SSWSYS013), and
the Strategic Priority Research Program of Chinese Academy of Sciences (Grant No. XDB34010000). 
Calculations have been performed in part at the HPC Cluster of KLTP/ITP-CAS and the Supercomputing Center,
Computer Network Information Center of CAS. 


\end{document}